\documentclass[%
 reprint,
 amsmath,amssymb,
 aps,
]{revtex4-2}

\usepackage{graphicx}
\usepackage{dcolumn}
\usepackage{bm}
\usepackage{comment}
\usepackage{footmisc}
\usepackage{hyperref}
\usepackage[mathlines]{lineno}

\begin{document}

\preprint{APS/123-QED}

\title{A Kuramoto Network in a Single Nonlinear Microelectromechanical Device}
\author{Samer Houri}
\email{samer.houri.dg@hco.ntt.co.jp}
  \affiliation{NTT Basic Research Laboratories, NTT Corporation,
3-1 Morinosato-Wakamiya, Atsugi-shi, Kanagawa 243-0198, Japan.}

\author{Motoki Asano}
\author{Hajime Okamoto}
\author{Hiroshi Yamaguchi}
  \affiliation{NTT Basic Research Laboratories, NTT Corporation,
3-1 Morinosato-Wakamiya, Atsugi-shi, Kanagawa 243-0198, Japan.}
\date{\today}

\begin{abstract}

This work presents a frequency 
multiplexed 3-limit cycles network in a multimode microelectromechanical nonlinear resonator. The network is composed of libration limit cycles 
and behaves in an analogous manner to a phase oscillator network. The libration limit cycles, being of low frequency, interact through the stress tuning of the resonator, and result in an all-to-all coupling that can be described by a Kuramoto model. Beyond the typically present cubic nonlinearity the modes in question do not require any special frequency ratios. Thus an interconnect free Kuramoto network is established within a single physical device without the need for electrical or optical coupling mechanisms between the individual elements.
\end{abstract}
\maketitle
\indent\indent\ \emph{Introduction.\textemdash}Oscillator networks, where a number of oscillators are coupled in such a way as to observe the emergence of a network-wide dynamics, are a topic of intense ongoing scientific and technological investigation \cite{strogatz2000kuramoto,strogatz2001exploring,strogatz2004sync}. Amongst the many possible constructions of networks, variants of the Kuramoto network \cite{kuramoto2003chemical,pikovsky2003synchronization,acebron2005kuramoto,dorfler2014synchronization} that is characterized by a sinusoidal coupling ansatz exhibit a wide range of phenomena such as synchronization \cite{dorfler2014synchronization}, phase transitions \cite{daido1992quasientrainment}, chimeras \cite{abrams2004chimera}, and spiral waves \cite{kim2004pattern}. The interest in Kuramoto networks extends from basic research to engineering applications that include neuromorphic \cite{romera2018vowel,hoppensteadt2001synchronization} and reservoir computation \cite{nakajima2020reservoir}.\\
\indent\indent\ Usually, an oscillator network, Kuramoto-type or otherwise, is experimentally constructed by establishing a number of individual (self-sustained) oscillators be they Josephson junctions \cite{wiesenfeld1996new}, spin-based \cite{romera2018vowel}, optical \cite{jang2018synchronization,mcmahon2016fully}, or electro-/opto-mechanical \cite{hoppensteadt2001synchronization,heinrich2011collective}; and then inducing coupling (controllable or otherwise) between these distinct elements. The coupling can be electrical \cite{matheny2019exotic,hoppensteadt2001synchronization}, optical\cite{safavi2014two,bagheri2013photonic,zhang2012synchronization}, or even mechanical \cite{shim2007synchronized,mahboob2016electromechanical}. This approach thus requires two main elements to construct a network, the oscillators (i.e., limit cycles) and the coupling.\\
\indent\indent\ In this work we construct a frequency multiplexed limit cycles network, where the different modes of the same structure form the oscillators, and the required coupling is generated via the nonlinear mode-coupling present in micro- and nano-electromechanical systems (M/NEMS). In nonlinear mode-coupling, which essentially is a non-resonant four-wave mixing, one mode would experience a frequency shift due to the added average strain that is generated by the amplitude of vibration of another mode \cite{westra2010nonlinear,matheny2013nonlinear,lulla2012nonlinear}. This average strain can be thought of as a (quasi)dc channel that transmits information about the amplitude of the different modes. Therefore, by relying on mode-coupling it is possible to reduce the essential elements of a network to the creation of the limit cycles only, and allow the mode-coupling to generate the necessary connections in a single physical device.\\
\indent\indent\ However, a straightforward implementation of a network in a multi-mode device with self-sustained oscillating modes is not possible without resorting to resonant four-wave mixing, i.e., internal resonance \cite{wang2021frequency,zhang20201,houri2020demonstration}. Internal resonance requires special frequency ratios that are difficult to scale to a large number of modes, and at the same time could result in strong coupling between the oscillating elements \cite{guttinger2017energy,czaplewski2018bifurcation,houri2019limit}, which is undesirable for an oscillator network.\\
\indent\indent\ To leverage the non-resonant nonlinear mode-coupling as a foundation for connecting frequency multiplexed limit-cycles in a network imposes two conditions. First, the mode-coupling transmits information about the amplitude and not the phase, thus a scheme is required to establish an amplitude-phase coupling so as to enable the structural mode-coupling to transmit the phase information between the different modes. Second, the information needs to be transmitted in a dc or quasi-dc fashion, since the non-resonant mode-coupling is a time average effect. Here, by quasi-dc we mean no more than a perturbation order frequency component, i.e. $\omega_{\textrm{quasi-dc}}\sim\mathcal{O(\epsilon)}$. Both of these requirements are fulfilled by using libration limit cycles \cite{houri2021librator}.\\
\indent\indent\ Self-sustained libration (or libration limit cycles or librators) are limit cycles taking place in the rotating frame of a harmonically driven resonator. More intuitively, it is possible to think of a traditional limit cycle phase oscillator as a circular trajectory in the laboratory frame phase-space. Whereas in a librator, the limit cycle is created around a harmonic drive with a force ($F_d$), and a frequency ($\omega_d$), such that the rotating frame phase space exhibits a limit cycle. This implies that these libration limit cycles, also approximated by a circular trajectory, are centered around the driven response of the system to $F_d$, and not centered around the origin of the rotating frame.\\
\indent\indent\ The off-center position of the libration limit cycle couples the phase and the amplitude, while the nature of the libration limit cycle is such that its frequency is equally a perturbation order frequency component, i.e, $\omega_L\sim\mathcal{O(\epsilon)}$ \cite{houri2021librator}. Thus librators fulfill the two criteria for coupling limit cycles through nonlinear mode coupling.\\
\indent\indent\ Therefore, whereas traditional networks involve a number of coupled nearly identical oscillators. This work studies networks that are formed by limit cycles that are nearly identical when considered in the rotating frame of their respective harmonic drives, and are coupled via the nonlinear mode-coupling.\\
\indent\indent\ The dynamics of the librator are not equivalent to those of a forced limit cycle oscillator, as the creation of the librator requires making the forced response itself unstable. To construct a librator requires a loop that feedbacks the rotating frame components (i.e., the quadratures), as shown schematically in Fig.~1(a). It is possible to think of a librator as a self-sustained amplitude modulation (AM), where the carrier frequency is set by the driving term ($F_d$, $\omega_d$), and the modulating frequency is the self-sustained libration frequency ($\omega_L$). Where $\omega_L$ depends on the linear and nonlinear parameters of the system.\\
\indent\indent\ The ability to synchronize librators to an externally injected tone has yet to be explored, let alone the ability to mutually synchronize coupled librators and establish a network. This is not necessarily as straightforward as synchronization of oscillators \cite{adler1946study,paciorek1965injection,barois2014frequency,matheny2014phase,houri2017direct}, since the librators are centered around varying frequencies (the modal frequencies) with no special frequency ratios. And the synchronization is to take place via the quasi-dc strain tuning around a quasi-dc frequency $\omega_{quasi-dc}$ as shown schematically in Figs.~1(b), (c). Therefore, we explore the ability to synchronize librators, first to an injected tone ($\omega_{sync}$), then to each other, before proceeding to construct a librator network.\\
\indent\indent\ \emph{Model.\textemdash}To account for the synchronizing influence, we modify the governing equation of the librator dynamics \cite{houri2021librator} to take on the following non-dimensional form\\
\begin{multline}
\ddot{x_i} + {(\gamma_i + \beta_i x_i^2)}\dot{x}_i + (1+\epsilon_{sync}\cos(\omega_{sync}t))x_i\\
+ \alpha_i x_i^3 = {{(F_{di} + f_i(t))}\cos(\omega_{di}t)}
\label{eqn:Eq1}
\end{multline}
where the subscript \emph{i} identifies the mode number. $x_i$ is the modal displacement, and $\gamma_i$, $\beta_i$, $\alpha_i$ are respectively the modal linear damping, nonlinear damping, and Duffing nonlinearity of the \emph{i}th mode (note that the equation is normalized so that the natural frequency $\omega_{0i}=1$). $\epsilon_{sync}$ and $\omega_{sync}$ denotes the magnitude and frequency of the quasi-dc frequency shift due to an externally applied synchronizing tone ($\omega_{sync}\ll1$). ${F_{di}}$ and $\omega_{di}$ are the amplitude and frequency of the  \emph{i}th mode driving force, and ${f_i(t)}$ is the modal feedback term that generates the libration limit cycle. We also introduce a detuning parameter $\delta_i$ such that $\omega_{di} = \omega_{0i}\times(1+\delta_i)$. Note that all the above parameters are considered to be perturbation order terms, i.e. $\gamma_i$, $\beta_i$, $\alpha_i$, $\epsilon_{sync}$, $\omega_{sync}$, $F_{di}$, $f_i(t)$, and $\delta_i$ $\sim\mathcal{O(\epsilon)}$.\\
\indent\indent\ Eq.~(1) is treated using the rotating frame approximation (RFA), whereby the displacement is expressed as $x_i = (A_ie^{iw_{di}t} + A_i^*e^{-iw_{di}t})/2$, and $A_i$ is the rotating frame complex amplitude of the \emph{i}th mode. This complex amplitude is in its turn decomposed into a static component and a dynamic (libration) component, denoted by $A_{0i}$ and $A_{Li}$ respectively. Where the static component $A_{0i}$ is the response of a standard Duffing resonator to a driving force ($F_{di}$, $ \omega_{di}$), and $A_{Li}$ is the libration limit cycle of the system in the rotating frame. $A_{Li}$ is considered to be centered around the static response $A_{0i}$.\\
\indent\indent\ The application of the RFA leads to a dynamical equation, that reads\\
\begin{multline}
\dot{A}_{Li}=-(i\delta_{Li}+\frac{1}{2}\gamma_{Li})A_{Li}+C_{Li}A^*_{Li}\\
+\frac{i}{2}\epsilon_{sync}\cos(\omega_{sync}t)A_{Li}
\label{eqn:Eq2}
\end{multline}
where $\delta_{Li}$ and $\gamma_{Li}$ are respectively the effective detuning and effective linear damping of the \emph{i}th mode libration motion ${A_{Li}}$. $C_{Li}$ is a complex constant that depends on the modal parameters, the driving force, and the feedback loop \cite{houri2021librator,suppinf}.\\
\indent\indent\ Note that since for the model in this work only the $A_{Li} \ll A_{0i}$ regime is being considered, we dropped all nonlinear terms in $A_{Li}$ from Eq.~(2), these include the terms that stabilize the magnitude of the limit cycle but play almost no role in setting its frequency (see supplementary materials for the derivation).\\
\indent\indent\ To explore the possibility of synchronization it is first necessary to presume a periodic oscillation for the libration motion, thus $A_{Li}$ is supposed to take on the following form $A_{Li} = B_{0i} + (B_{i}e^{i\omega_{Li}t} + B_{i}^*e^{-i\omega_{Li}t})/2$, where $B_{i}$ is the periodic component of the libration motion, and $B_{0i}$ is a non-zero static component that results from the non-symmetric shape of the libration orbit, knowing that $B_{0i}=U_{0i}e^{i\phi_{0i}}$, and $B_{i}=U_ie^{i\phi_i}$.  For convenience, we also express the steady state driven response as $A_{0i}=R_{0i}e^{i\theta_{0i}}$. In essence, the introduction of $B_{Li}$ and $B_{0i}$, amounts to a second rotating frame approximation, or a second order perturbation analysis, meaning the time scales associated with these dynamics are $\sim\mathcal{O}(\epsilon^{-2})$.\\
\indent\indent\ By inserting the expansion of $A_{Li}$ into Eq.~(2), developing, and collecting the relevant terms (see supplementary materials for details), it is possible to obtain the following Adler-like \cite{adler1946study,paciorek1965injection,pikovsky2003synchronization} phase locking equation\\
\begin{equation}
    {\dot{\Phi}_{Li} = {\Omega}_{i} + {P}_{i}\sin({\Phi}_{Li})}
    \label{eqn:Eq3}
\end{equation}
where ${\Omega}_{i}$ is an effective detuning parameter between the \emph{i}th mode libration frequency and the synchronizing influence \cite{suppinf}, ${P}_{i}$ is the synchronization forcing, and ${\Phi}_{Li}$ represents an effective phase difference between the libration limit cycle and the synchronizing influence.\\
\indent\indent\ For synchronization to take place requires that $\dot\Phi_{Li}=0$, while the terms ${\Omega}_{i}$, ${P}_{i}$, and ${\Phi}_{Li}$ need to be derived from the governing equations. These terms differ depending on whether synchronization is a result of an external forcing or mutual synchronization between interacting libration limit cycles.\\
\indent\indent\ \emph{Injection locking.\textemdash}First, we consider the synchronizing effect of an applied external force where two cases will be treated, these are $\omega_{sync}\approx\omega_{Li}$, and $\omega_{sync}\approx 2\omega_{Li}$. By retracing the steps for each one of these cases we derive expressions for the forcing parameter in Eq.~(3), which for the former case takes the form ${P}_i(\omega_{sync}\approx\omega_{Li}) = (\epsilon_{sync}/4)\times(U_{0i}/U_i)$, and for the latter case is ${P}_i(\omega_{sync}\approx2\omega_{Li}) = \epsilon_{sync}/4$.\\
\indent\indent\ The locking range is calculated by setting $\dot{\Phi}_{Li }= 0$ in Eq.~(3), which gives a simple $\pm\epsilon_{sync}=4\Omega_i$ relation for the case of frequency locking with $\omega_{sync}\approx2\omega_{Li}$. The case of $\omega_{sync}\approx\omega_{Li}$ is difficult to calculate analytically, as the locking range depends on the asymmetry of the orbit $U_{0i}$, which can only be determined by numerically integrating the governing ODE. However, it is possible to set an upper bound of $U_{0i} = U_{i}$. Thus, $\pm\epsilon_{sync}=4\Omega_i$ can be considered to be a bound on both synchronization scenarios.\\
\indent\indent\ We experimentally investigate the synchronization dynamics of librator limit cycles using a piezoelectrically actuated GaAs MEMS clamped-clamped beam device that is 100 $\mu$m in length, 20 $\mu$m wide, and 600 nm in thickness, see \cite{yamaguchi2017gaas} for more information on device fabrication. The device is placed in a vacuum chamber with a pressure of $\sim$ 1 mPa \cite{suppinf}, excited electrically, and its vibrations are measured optically using a laser Doppler vibrometer (LDV). A dc voltage component ($\sim$ -1 V) is constantly applied, which ensures that the actuation remains linear by avoiding the Schotkey behavior of the metal-semiconductor junction \cite{yamaguchi2017gaas,houri2019limit}. In addition, the applied dc component results in a constant strain in the structure that shifts the resonance frequency, this electrically controllable frequency tuning plays the crucial role of generating $\epsilon_{sync}$ in Eq.~(1) by modulating the dc voltage component.\\
\indent\indent\ The device in question possesses two electrodes on each end of the structure length, see representation in Fig.~1(a), by equally actuating both electrodes (not shown in schematic) only odd modes are efficiently excited \cite{houri2020demonstration}. Three odd modes are accessible these are the first, third, and fifth out-of-plane flexural modes, respectively, with the following modal frequencies $\omega_{01} = 2\pi\times321$ kHz, $\omega_{03} = 2\pi\times954$ kHz, $\omega_{05} = 2\pi\times2.325$ MHz. These modes equally exhibit a Duffing-type nonlinearity, Fig.~2(a), and nonlinear damping.\\
\indent\indent\ We generate libration limit cycles using feedback loops that are functionally equivalent to the ones shown in Fig.~1(a) (see supplementary materials for details on experimental setups). Once the limit cycles are established, we sweep the drive frequencies $\omega_{di}$ (i.e., $\delta_{di}$) and quantify the libration frequencies, i.e., $\omega_{Li}$, which are then plotted in Fig.~2(b).\\
\indent\indent\ To study synchronization due to an external forcing, we choose the case of zero detuning, i.e., $\delta_{di} = 0$, and apply a weak tone on top of the dc bias. Naturally, when investigating external synchronization only one librator feedback loop is active at a time so as not to have mutually interacting librators.\\
\indent\indent\ Since the injected signal frequency is on the order of the librator frequency and hence much smaller than the modal frequencies, i.e. $\omega_{sync} \sim \omega_{Li} \ll \omega_{di}$, its effect is to modulate the resonance frequencies of the structure, by modulating the dc bias, rather than to directly force the resonant modes, as depicted schematically in Fig.~1(b). Experimentally, the effect of the injected signal on the librator limit cycles is clearly visible in Fig.~2(c), where the phase space plot shows the locking of the trajectories to the phase of the outside signals.\\
\indent\indent\ Thereafter, the frequency of the injected signal is swept producing a noticeable locking interval which is typically seen and expected from phase-locked systems \cite{pikovsky2003synchronization,paciorek1965injection,matheny2014phase}, see Fig.~2(d). Subsequently, a 2-dimensional parameter sweep is undertaken, where the force ($\epsilon_{sync}$) and frequency ($\omega_{sync}$) of the locking signal are swept, these 2D sweeps are shown in Fig.~2(e), for both the $\omega_{sync} \approx \omega_{Li}$ case and for the $\omega_{sync} \approx 2\omega_{Li}$ case. These sweeps demonstrate locking regions similar to Arnold tongues.\\
\indent\indent\ These plots provide reassuring evidence of the validity of the perturbation analysis and the resulting Eq.~(3), since by having plotted the detuning and forcing in normalized terms, we find that the boundary of the synchronization intervals are reasonably well delineated by the linear relation $\epsilon_{sync} = \pm 4\Omega_i$ as predicted by the model. On a side note, it is interesting to remark that in absolute terms the frequency locking ratio, defined as $= \omega_{di}/\omega_{sync}$, is on the order of 1000.\\ 
\indent\indent\ \emph{Mutual synchronization.\textemdash}Having established the potential of librator limit cycles to phase-lock to an external source, we now investigate the ability of multiple librator limit cycles, each centered around a different mode, to interact and synchronize. In this case, structural mode-coupling that is present in nonlinear M/NEMS devices \cite{westra2010nonlinear,matheny2013nonlinear,lulla2012nonlinear} acts as a stress tuning mechanism that provides a low frequency coupling channel between the librator limit cycles, see Fig.~1(c). Only quasi-dc mode-coupling is being considered, which implies that no resonant four-wave mixing should exist between the modes, i.e. $2\omega_i - \omega_j - \omega_k \neq 0$ \cite{houri2019modal,kurosu2020mechanical}.\\
\indent\indent\ For mutual librator synchronization the parametric frequency tuning term in Eq.~(1) is replaced by the standard mode-coupling terms \cite{westra2010nonlinear,matheny2013nonlinear,lulla2012nonlinear}, i.e. $\big(\epsilon_{ij}x_j^2 + \epsilon_{ik}x_k^2 + ...\big)x_i$, where $\epsilon_{ij}$, $\epsilon_{ik}\cdots$ are the mode-coupling constants between the $i$th mode and modes $j, k \cdots$. In the rotating frame of the $i$th mode, the mode-coupling terms reduce to $\frac{i}{4}\epsilon_{ij}\mid A_j\mid^2A_i + \frac{i}{4}\epsilon_{ik}\mid A_k\mid ^2A_i+\cdots$ \cite{westra2010nonlinear}.\\
\indent\indent\ However, the modal amplitudes are no longer constant, as they are slowly modulated by the libration terms. Therefore, they can be written as $\mid A_j\mid=\mid A_{0j} + A_{Lj}\mid$, and $\mid A_k\mid=\mid A_{0k} + A_{Lk}\mid, \cdots$. By placing the modulated amplitudes in the mode-coupling terms and developing, we obtain the following phase relationship (see supplementary materials for derivation \cite{suppinf})\\
\begin{equation}
    {\dot{\phi}_i = {\Omega}_{i}+\sum_{j} k_{ij}\sin(\Delta\phi_{ij}+\psi_{0j})}
    \label{eqn:Eq4}
\end{equation}
where $j$ denotes all the modes that couple to the $i$th mode, $\Omega_i$ is an effective detuning parameter, and $\Delta\phi_{ij}$ represents the phase difference between the two limit cycles, i.e. $\Delta\phi_{ij} = \phi_j-\phi_i$. The constants $k_{ij}$ and $\psi_{0j}$ are rather involved amplitude and phase parameters (see supplementary material \cite{suppinf}) that depend on the $j$th modal amplitude, mode-coupling, and other parameters. Note that $\mid k_{ij}\mid > 0$, since mode-coupling can not be turned off.\\
\indent\indent\ It is significant that Eq.~(4) corresponds to a variant of the well-known Kuramoto model \cite{sakaguchi1986soluble,acebron2005kuramoto,kuramoto2003chemical,daido1992quasientrainment,dorfler2014synchronization}. Thus, the mode-coupling mechanics naturally give rise to a Kuramoto-type network in a multimode M/NEMS device, when librator limit cycles are excited around these modes.\\
\indent\indent\ In order to investigate the behaviour of mulimode librator networks, we first start with the simplified case of only two coupled librators $i$ and $j$. Thus, only two phase equations are necessary, those of $\dot{\phi}_i$ and $\dot{\phi}_j$. By considering only the difference between the two equations, i.e., $\Delta\dot{\phi}_{ij} = \dot{\phi}_i - \dot{\phi}_j$, Eq.~(4) can be rewritten as\\
\begin{equation}
    {\Delta\dot{\phi}_{ij} = {\Delta\Omega}_{ij}+\mu\sin(\Delta\phi_{ij})+\nu\cos(\Delta\phi_{ij})}
    \label{eqn:Eq}
\end{equation}
where $\Delta\Omega_{ij}$ is the frequency difference between the two free librators, and $\mu$ and $\nu$ are rearranged constants (see supplementary materials \cite{suppinf}).\\
\indent\indent\ Equation~(5) is superficially different from the classical two-coupled phase oscillators equation \cite{matheny2014phase,pikovsky2003synchronization,aronson1990amplitude,wirkus2002dynamics,weiss2016noise} (with constant amplitude). The difference is due to the asymmetry in the coupling and the presence of the phase components $\psi_{0j}$, $\psi_{0i}$. Nevertheless, the fact remains, that Eq.~(5) has only two possible outcomes, either the librators synchronize or they do not.\\
\indent\indent\ Experimentally, we investigate this regime by generating two libration limit cycles around the modes of interest. Since the libration limit cycles frequencies depend, amongst other things, on the force detuning term $\delta_i$ then by sweeping the latter the libration frequencies are adjusted until nearing a 1:1 ratio, upon which they should lock.\\
\indent\indent\ An experimental example of synchronization between the mode 3 librator and mode 5 librator is shown in Fig.~3(a). In the figure, the drive frequency of mode 5 (and  hence the libration frequency $\omega_{L5}$) is left unchanged, while the detuning term of mode 3 is swept. The extracted frequency difference and frequency ratio are plotted, where it is easy to identify the 1:1 locking range between the two librators. Interestingly, a miniature plateau corresponding to a 2:1 locking ratio is equally observed.\\
\indent\indent\ The effect of frequency locking between the librators is also shown in the insets of Fig.~3(a), by tracing the in-phase components versus each other, i.e. $X_3$ vs. $X_5$, where for a 1:1 locking ratio a simple circle is formed, for a 2:1 locking ratio a figure 8 is formed, and for unlocked librators the trace would simply fill a rectangle (not shown).\\ \indent\indent\ By expanding these mutual synchronization measurements to 2-dimensional sweeps, where the drive frequencies of both modes are swept, and then plotting the resulting phase difference $\Delta\phi_{ij}$, new features become apparent, as shown in Fig.~3(b). For one the 1:1 synchronization region is quite visible for all pairs of librators. Furthermore, higher order locking, e.g. 1:2, 2:1, 3:1, and 4:1, regions can be identified. These higher order locking regions are not directly predicted by Eq.~(5) since the inclusion of higher order terms would be necessary to account for them.\\
\indent\indent\ The results shown in Fig.~3 represent an effective confirmation of Eq.~(5), through the demonstration of pairwise synchronization. Yet, this confirmation remains qualitative, since there are simply too many free parameters hidden in the terms $\Omega_i$, $\psi_{0j}$ and $k_{ij}$, in Eq.~(4), which prevents the possibility of having an approximate quantitative bound as was done for Eq.~(3).\\
\indent\indent\ An undesirable effect is equally visible in Fig.~3(b), which is due to the fact that the resonance frequency ratio of modes 1 and 3 is $\omega_3/\omega_1 \approx 3$, thus a region of resonant energy transfer, or internal resonance, can be accessed \cite{houri2019limit,houri2020demonstration}. In this work, such effect is undesirable, as it changes drastically the nature of the coupling, and the region where this resonant energy transfer takes place is avoided.\\
\indent\indent\ \emph{Network.\textemdash}If the number of librators is increased further then the dynamics changes significantly as $N\geqslant 3$, since with 3 or more nodes the system becomes a network, see Fig.~4(a), with the possibility of partial synchronization and more complex states \cite{mendelowitz2009dynamics,bridge2009dynamics,rompala2007dynamics,wojewoda2016smallest,maistrenko2017smallest,martens2010chimeras,matheny2019exotic,bhaskar2021synchronization,martens2013chimera,ashwin2015weak,hart2016experimental,acebron2005kuramoto,dorfler2014synchronization,rohm2016small,dudkowski2020small,arenas2008synchronization,lauter2015pattern,lauter2017kardar}.\\ 
\indent\indent\ We proceed to study the 3-node network dynamics by activating all three feedback loops simultaneously, and changing the dc from -1V to -1.5V while accounting for the slight shift in resonance frequencies. Seeing the substantial size of the parameter space, an exhaustive, or even systematic, sweep is impractical. Instead a linear search procedure is used to adjust the values of the detunings (i.e., $\delta_1$, $\delta_3$, $\delta_5$), whereas all other experimental parameters are kept constant. The linear search has for objective the minimization of the libration frequencies spread.\\
\indent\indent\ As the libration limit cycles frequencies are brought closer together, they transition from an unsynchronized state, to a partially synchronized state (2 modes synchronized), to a fully synchronized state, depending on their respective detunings. These states can be seen by plotting the modes' $X$-quadratures against each other in a 3D plot, as shown in Fig.~4(b). The effect of partial and full synchronization on the phase space trajectory is clearly observed, where the trajectory moves from filling a volume (unsynchronized), to the surface of a cylinder (partial synchronization), to a simple ring (full synchronization). \\
\indent\indent\ Once the synchronization parameters are established, we explore a small volume in the $\delta_1$, $\delta_3$, $\delta_5$ space around those parameters. To quantify the degree of synchronization we use the time average of the Kuramoto order parameter $\bar{r}$, where $r(t)e^{i\Psi(t)} = \sum_{i} e^{i\phi_i(t)}$ \cite{kuramoto2003chemical,acebron2005kuramoto,schroder2017universal}. The results are shown in the 2D plot in Fig.~4(c). The 2D sweep shows regions of synchronization, identified by the bright color area. Surprisingly, the value of $\bar{r}$ varies roughly between 0.4 for the unsynchronized case, and 0.7 for the synchronized case, whereas it should vary between 0 and 1 for those two cases, respectively. This is likely due to the presence of higher order libration terms as implied from the plots of the mean field ($r(t)$, $\Psi(t)$) in Fig.~4(d). The mean field traces an elliptical trajectory unlike the traditional mean field representation on a circle. For comparison, the mean field of the unsynchronized cases are also shown in Fig.~4(d), unsurprisingly they show no pattern.\\
\indent\indent\ The parameters $\Omega_i$, $k_{ij}$, and $\psi_{0j}$ can all be manipulated experimentally (to some extent) by changing the drive forces' detuning and magnitude ($\delta_i$, $F_{di}$) as well as the phase of the feedback loops ($\Theta_i$). The presence of the phase term $\psi_{0j}$ in Eq.~4 could lead to frustration in the system \cite{panaggio2015chimera,sakaguchi1986soluble,daido1992quasientrainment}. This was tentatively observed in this work, where a $\pi/2$ phase shift on the lock-in amplifier of the fifth mode resulted in the librators unable to synchronize, even after the linear search algorithm brought their frequencies to be practically overlapping. These controllable parameters therefore provide a valuable means to experimentally tailor the properties of the network to be studied.\\
\indent\indent \emph{Conclusions.\textemdash}To summarize, this work builds on the recently introduced MEMS librator to demonstrate the potential of the librator limit cycles to be synchronized to an outside force as well as to each other. This latter effect is mediated through the structural mode-coupling, where the libration motion, being of low frequency, couples through the stress-tuning of the structure. The emergent network thus formed is best described by a Kuramoto model, despite the various limit cycles and their collective mean field taking place at largely different frequencies. This work therefore dispenses with the need for electrical or optical coupling mechanisms as well as provides experimental means to control the network properties.\\
\indent\indent\ With a more streamlined experimental setup, e.g., replacing the lock-ins with RF power detectors, it would be possible to scale the number of nodes using simple of-the-shelf MEMS devices and control electronics. Furthermore, the principles described in this work are equally applicable to other types of systems with Kerr-type nonlinearity, like optical resonators and microwave cavities.\\
\begin{figure}[th]
	\graphicspath{{Figures/}}
	\includegraphics[width=85mm]{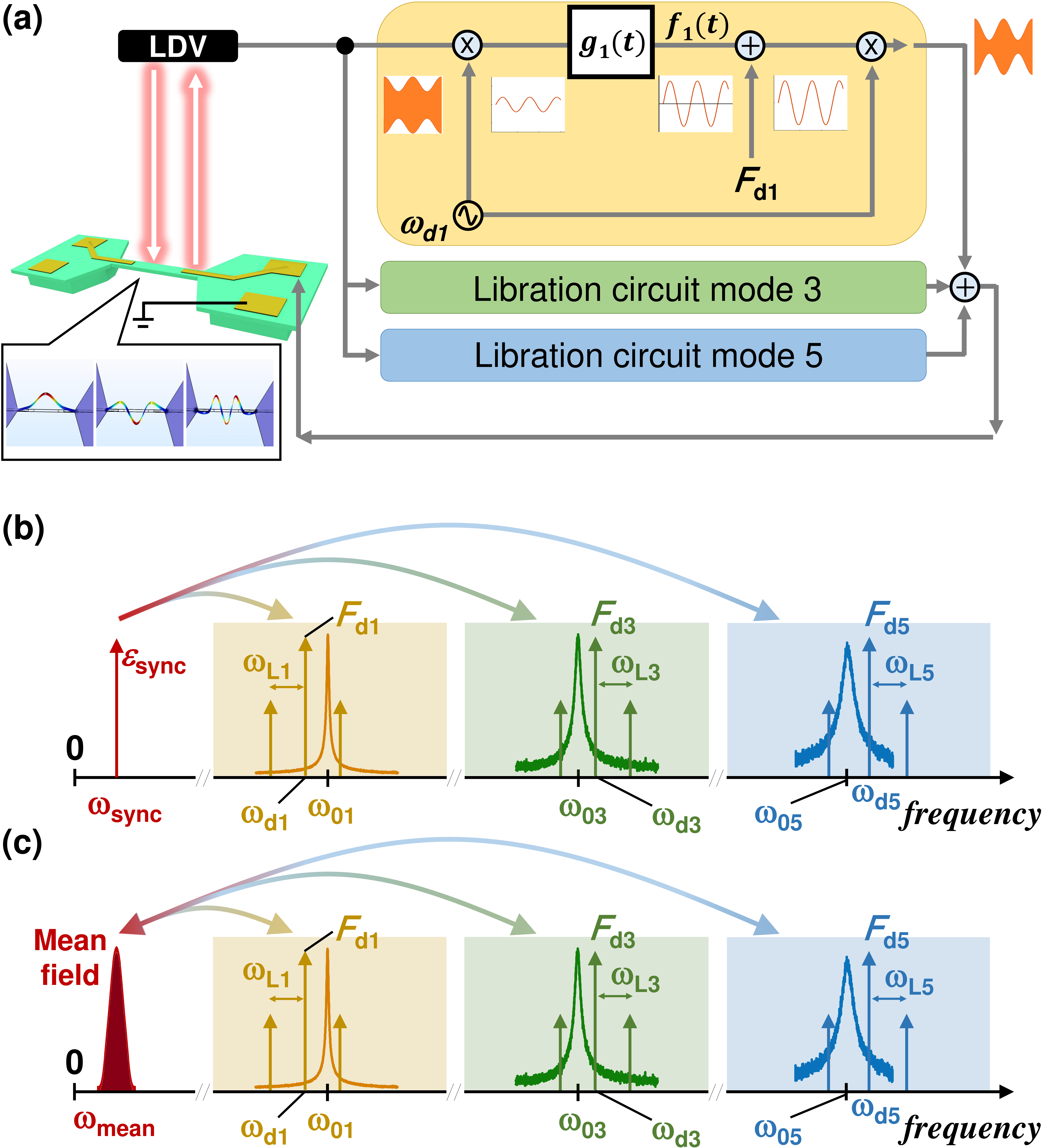}
	\caption{(a) Schematic representation of the experimental setup. The output from a LDV is passed through 3 lock-in amplifiers, each set to one of the drive tones (details shown only for one loop). The output from the lock-ins, each representing an individual mode, are high-pass filtered and passed through gain circuits ($g_{1,2,3}(t)$) then added to the drive force ($F_{d1,d2,d3}$) and up-converted to their respective drive frequencies, combined, and used to excite the MEMS device. A signal trace of mode 1 is shown for clarification. The lock-ins outputs are sampled by a digital oscilloscope (not shown).
	(b) Schematic representation of the stress tuning-mediated forced synchronization due to an injected tone ($\epsilon_{sync}$) around $\omega_{sync}\sim \omega_{Li}$. (c) Schematic representation of mode coupling-mediated interactions. The slow amplitude modulation, resulting from the libration motion of a frequency of $\omega_{L1,L2,L3}$ causes a quasi-dc stress tuning (mean field) that couples between the respective modes.}
\end{figure}

\begin{figure}[th]
	\graphicspath{{Figures/}}
	\includegraphics[width=85mm]{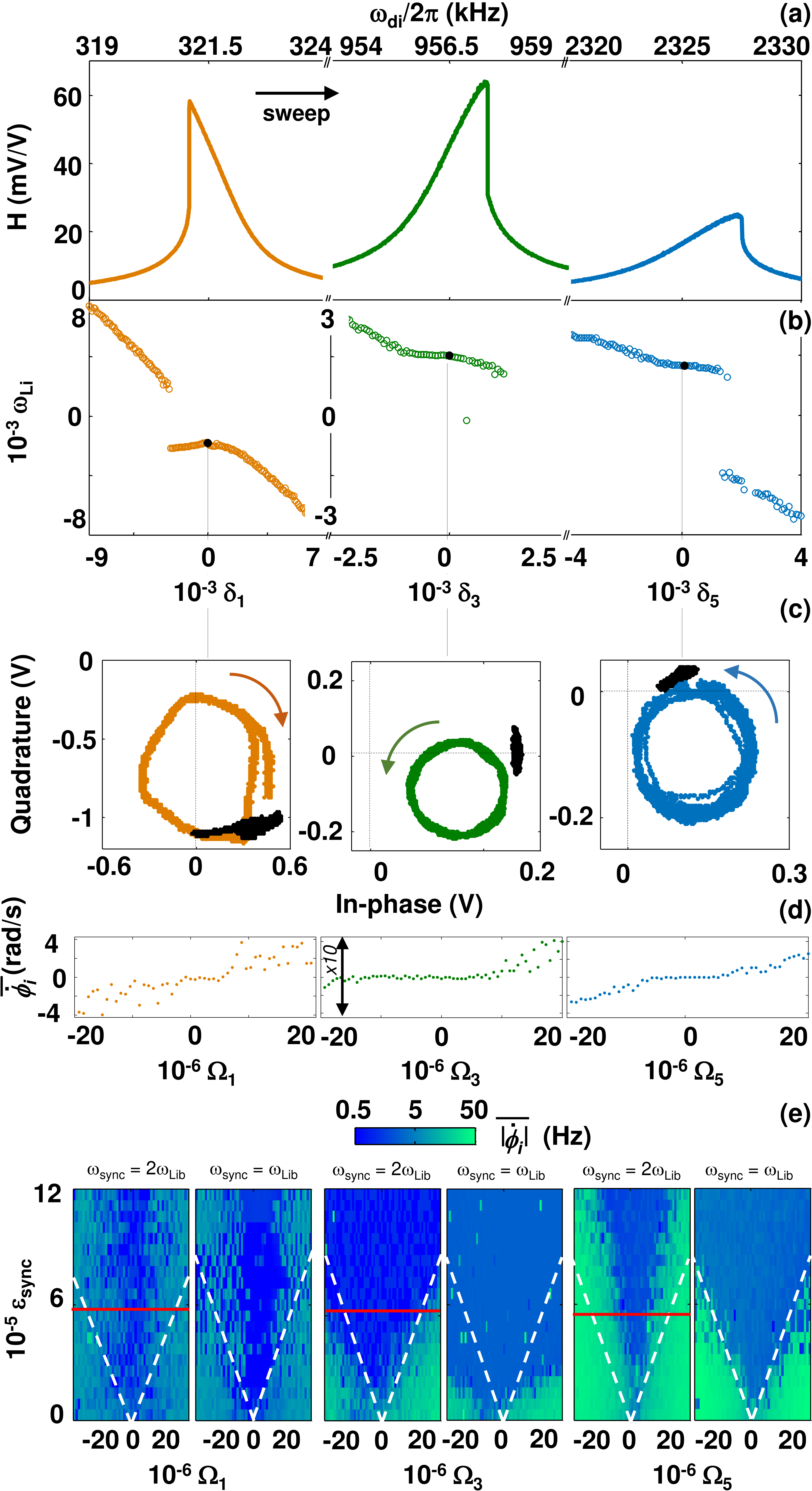}
	\caption{Injection locking of libration limit cycles. (a) Frequency response of the various modes for drive amplitudes of, 500 mV, 300 mV, and 1 V for the first (yellow, left), third (green, center), and fifth (blue, right) modes, respectively. 
	(b) Experimentally obtained normalized libration frequencies plotted as a function of drive detuning ($\omega_{di}$, $\delta_i$). The operating points for investigating synchronization are selected for zero detuning (i.e. $\delta_{i}=0$), points shown in black. The sign of the libration frequencies indicates the direction of libration. (c) The experimental data showing the difference between free running (colored) and locked (black) libration limit cycles. The dotted lines indicate the location of the axis of the phase space plane. (d) Locking range obtained for $F_{sync}=5.8\times10^{-6}$ (the vertical scale is multiplied by 10 for the 3rd and the 5th modes). (e) The measured synchronization tongues for $\omega_{sync}\approx2\omega_{Li}$ (left side panels) and for $\omega_{sync}\approx\omega_{Li}$ (right side panels) for the first, third, and fifth modes respectively. The dashed white lines delineate the maximum locking range area, i.e. $\epsilon_{sync}=4\Omega_i$, as obtained from Eq.~(3). Blue and green, indicate a locked and a running phase, respectively. The red lines indicate the data traces in (d).}
\end{figure}

\begin{figure}[th]
	\graphicspath{{Figures/}}
	\includegraphics[width=85mm]{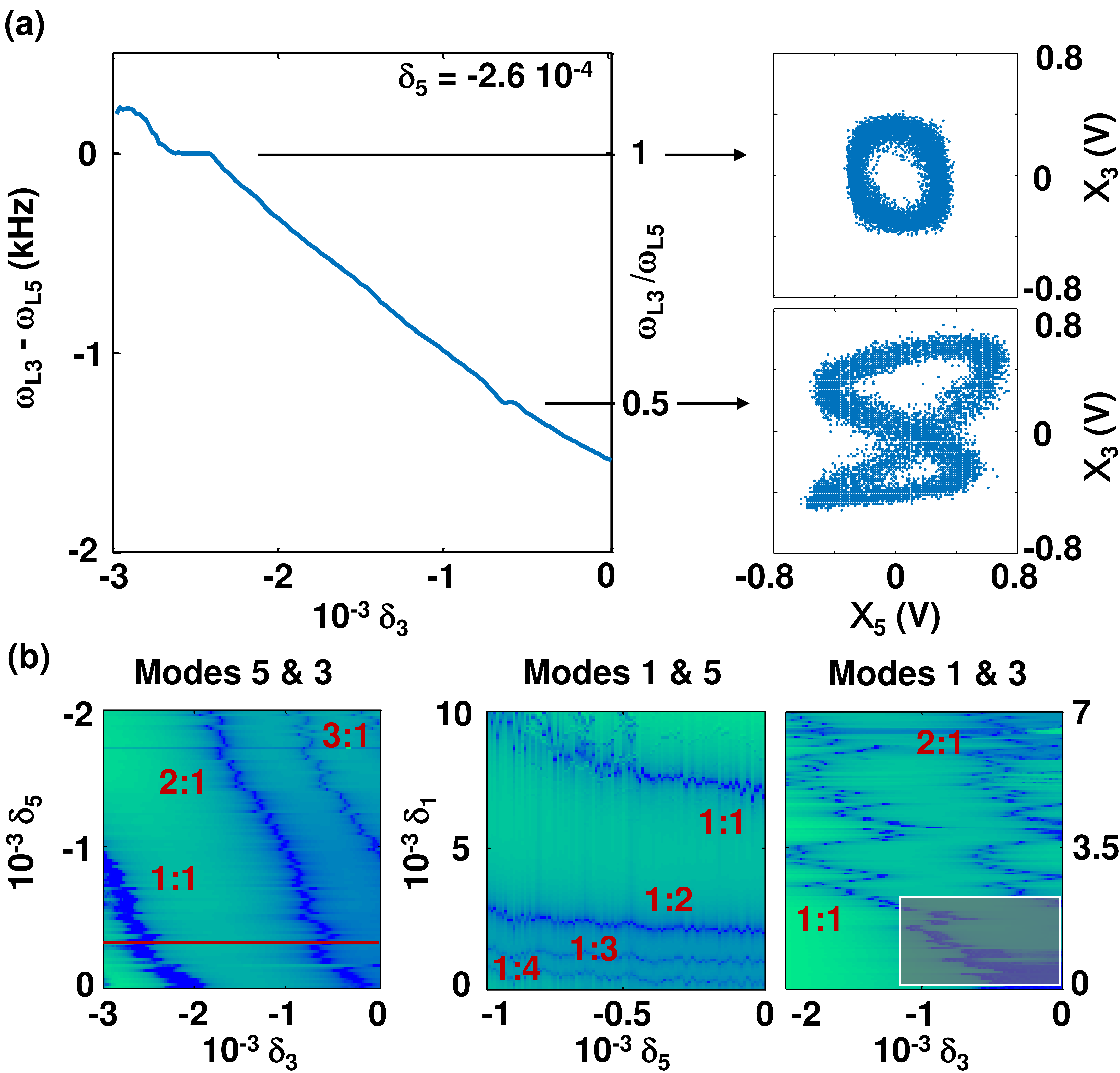}
	\caption{ Pairwise mutual synchronization of librators. (a) Modes 3 and 5, the main plot (left) shows the frequency difference (left vertical axis) and the frequency ratio (right vertical axis) between the two librator limit cycles, plotted as a function of the swept detuning of mode 3, while the detuning of mode 5 is kept constant ($\delta_{5} = -0.26\times10^{-3}$). The 1:1 locking plateau is clearly visible, a small 2:1 plateau is equally detectable. Right inset, the phase space plots showing the AC components of the quadratures of the librator limit cycles plotted against each other, for the case of 1:1 (top) and 2:1 (bottom) synchronization. (b) 2-D sweeps of the detunings for the pairs 3 $\&$ 5 (left), 1 $\&$ 5 (center), and 1 $\&$ 3 (right), showing the synchronization regions in blue. In addition to the clearly visible 1:1 synchronization region, higher order synchronization are also identifiable. The red line in the leftmost figure corresponds to the plot in (a). The shaded area in the right panel indicates the region where resonant energy transfer takes place. The color scale is in arbitrary units and optimized for visibility.}
\end{figure}

\begin{figure}[th]
	\graphicspath{{Figures/}}
	\includegraphics[width=85mm]{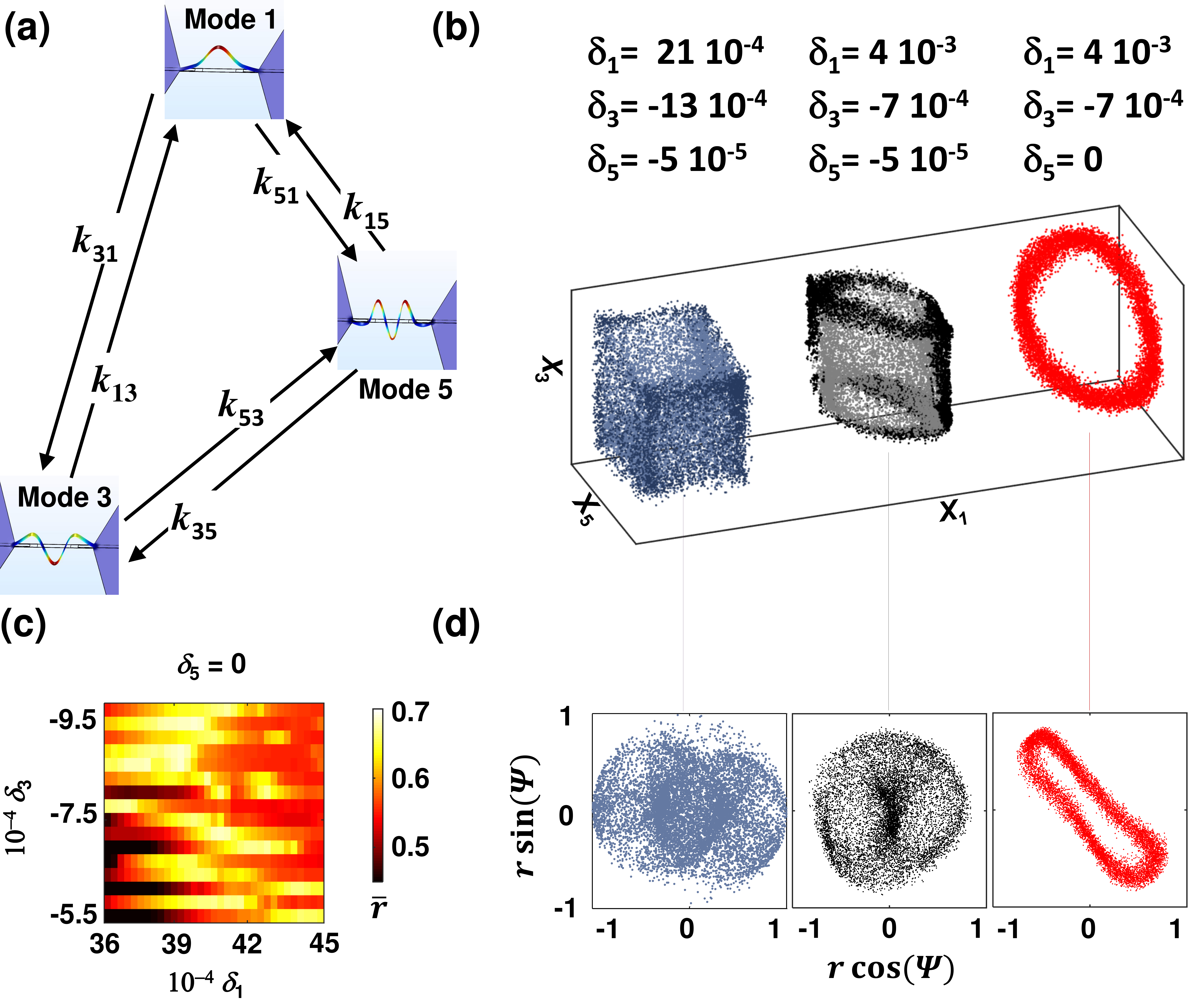}
	\caption{(a) Schematic representation of the 3-mode network and the asymmetric coupling. (b) Plots of the experimentally obtained $X$-quadratures for the unsynchronized case (blue points), partially synchronized case (black points), and fully synchronized case (red points). The plots are in arbitrary units, the edges of the cube and cylinder are emphasised for clarity, and the respective detunings of the data sets are noted on top. (c) 2D plot of the $\bar{r}$ parameter as a function of $\delta_1$ and $\delta_3$ ($\delta_5 = 0$), the color bar indicates that the value of $\bar{r}$ fluctuates between 0.4 and 0.7. (d) The mean field trajectory for the unsynchronized, partially- and the fully- synchronized cases shown in (b), blue, black , and red traces respectively. The fully synchronized case shows a closed trajectory that is however not circular, whereas the unsynchronized and partially synchronized case do not show any particular trajectory.}
\end{figure}

\bibliography{apssamp}

\end{document}